\documentclass[preprint,aps,showpacs,draft]{revtex4}
\usepackage{epsf}
\usepackage{dcolumn}% Align table columns on decimal point
\usepackage{bm}% bold math
\everymath{\displaystyle}
\begin{document}

\title{Tensor electric polarizability of the deuteron in storage-ring experiments}

\author{Alexander J. Silenko}
\email{silenko@inp.minsk.by}

\affiliation{Institute of Nuclear Problems, Belarusian State
University, Minsk 220080, Belarus}

\date{\today}

\begin {abstract}
The tensor electric polarizability of the deuteron gives important
information about spin-dependent nuclear forces. If a resonant
horizontal electric field acts on a deuteron beam circulating into
a storage ring, the tensor electric polarizability stimulates the
buildup of the vertical polarization of the deuteron (the
Baryshevsky effect). General formulas describing this effect have
been derived. Calculated formulas agree with the earlier obtained
results. The problem of the influence of tensor electric
polarizability on spin dynamics in such a deuteron
electric-dipole-moment experiment in storage rings has been
investigated. Doubling the resonant frequency used in this
experiment dramatically amplifies the Baryshevsky effect and
provides the opportunity to make high-precision measurements of
the deuteron's tensor electric polarizability.
\end{abstract}
\pacs {21.45.+v, 11.10.Ef, 21.10.Ky}

\maketitle

\section{Introduction}

Electric and magnetic polarizabilities are important properties of
deuteron and other nuclei. Tensor electric and magnetic
polarizabilities are defined by spin interactions of nucleons. In
particular, measurement of the tensor electric polarizability of
deuteron gives an important information about an interaction
between spins of nucleons and provides a good possibility to
examine the theory of spin-dependent nuclear forces.

The method of determination of this important electromagnetic
property of deuteron has been proposed by V. Baryshevsky {\it et
al.} \cite{Bar1,Bar3,Bar4}. If an electric field acts on a
deuteron beam circulating into a storage ring, the presence of the
tensor electric polarizability leads to the appearance of an
interaction quadratic in spin. When the electric field in the
particle's rest frame oscillates at the resonant frequency, the
effect similar to the nuclear magnetic resonance (NMR) takes
place. This effect stimulates the buildup of the vertical
polarization (BVP) of deuteron beam \cite{Bar1,Bar3,Bar4}.

In the present work, we derive general formulae describing the BVP
caused by the tensor electric polarizability of deuteron in
storage rings (the Baryshevsky effect).
Another effect defined by the tensor magnetic polarizability of
deuteron consists in the spin rotation in the horizontal plane at
two frequencies instead of expected rotation at the g$-$2
frequency \cite{Bar1,Bar3,Bar4}. In the above cited works, the
approach based on equations defining dynamics of polarization
vector and polarization tensor has been used. To check obtained
results and develop a more general theory, we follow the quite
different method of spin amplitudes (see Refs. \cite{CR,MZ}). In
the present work, this method is partially changed. We use the
matrix Hamiltonian for determining an evolution of %the
spin wave function.

The Baryshevsky effect should be taken into account at a search
for the electric dipole moment (EDM) of deuteron
\cite{Bar1,Bar3,Bar4}. An existence of the electric dipole moment
of deuteron also leads to the BVP. It is planned to measure the
BVP in the deuteron EDM experiment in storage rings
\cite{YOr,YS,OMS}. Since the Baryshevsky effect can imitate the
existence of the EDM, the spin dynamics caused by this effect
should be investigated in detail.

We use the relativistic system of units $\hbar=c=1$.

In the next section, we review the main aspects of the used
Hamiltonian approach in the method of spin amplitudes. In Section
III, we briefly discuss the form of the Hamilton operator in a
cylindrical coordinate system. Section IV is devoted to the
calculation of corrections to the Hamilton operator for the tensor
polarizabilities of deuteron. A solution of the matrix Hamilton
equation is given in Section V. In Section VI, we calculate the
spin dynamics expressed by the evolution of the vertical component
of polarization vector. A detailed analysis of new experimental
possibilities for measuring the tensor electric polarizability of
deuteron is performed in Section VII. Section VIII is dedicated to
the differentiation of effects of the EDM and the tensor electric
polarizability in the deuteron EDM experiment. Finally, in Section
IX we discuss previously obtained formulae and summarize the main
results of the work.

\section{Hamiltonian Approach in the Method of Spin Amplitudes}

The method of spin amplitudes uses quantum mechanics formalism to
more easily describe spin dynamics (see Refs. \cite{CR,MZ}). For
spin-1/2 particles, it is mathematically advantageous to use this
formalism because transporting the two-component spin wave
function (spinor) $\Psi$ is simpler than transporting the
three-dimensional polarization vector $\bm P$. The relationship
between them is given by the expectation value of the Pauli spin
vector $\bm\sigma$:
\begin{eqnarray} \bm
P=\Psi^\dag\bm\sigma\Psi,~~~ \Psi=\left(\begin{array}{c}
C_{+1/2}(t)
 \\ C_{-1/2}(t) \end{array}\right), \label{eqhlf}\end{eqnarray} where
$C_{+1/2}(t)$ and $C_{-1/2}(t)$ are the time-dependent amplitudes.
Together with the identity matrix the Pauli matrices generate an
irreducible representation of the SU(2) group.

Algebraically, the SU(2) group is a double cover of the
three-dimensional rotation group SO(3). Therefore, the formalism
based on the Pauli matrices is applicable to particles/nuclei with
arbitrary spin if an effect of spin rotation is analyzed. The spin
rotation can also be exhaustively described with the polarization
vector $\bm P$ which is defined by
\begin{eqnarray}
P_i =\frac{<S_i>}{S}, ~~~ i,j=x,y,z, \label{eq1P}\end{eqnarray}
where $S_i$ are corresponding spin matrices and $S$ is the spin
quantum number. The polarization vector being an average spin is a
strictly classical quantity (see Ref. \cite{Lee}) whose evolution
can be investigated in the framework of classical spin physics.

Particles with spin $S\geq1$ also possess a tensor polarization.
Main characteristics of such a polarization are specified by the
polarization tensor $P_{ij}$ which is given by \cite{MShY}
\begin{eqnarray}
P_{ij} = \frac{3 <S_iS_j + S_jS_i>-2S(S+1)\delta_{ij}}{2S(2S - 1)},
~~~ i,j=x,y,z. \label{eq1}\end{eqnarray} The polarization tensor
satisfies the conditions $P_{ij}=P_{ji}$ and
$P_{xx}+P_{yy}+P_{zz}=1$ and therefore has five independent
components. Additional tensors composed of products of three or more
spin matrices are needed only for the exhaustive description of
polarization of particles/nuclei with spin $S\ge3/2$.

The spin matrices for spin-1 particles have the form
\begin{equation}
\begin{array}{c}
S_x=\frac{1}{\sqrt{2}}\left(\begin{array}{ccc} 0 &
 1 & 0\\ 1 & 0 & 1\\ 0 & 1 & 0
 \end{array}\right), ~~ S_y=\frac{i}{\sqrt{2}}\left(\begin{array}{ccc} 0 &
 -1 & 0\\ 1 & 0 & -1\\ 0 & 1 & 0
 \end{array}\right), ~~ S_z=\left(\begin{array}{ccc} 1 &
 0 & 0\\ 0 & 0 & 0\\ 0 & 0 & -1
 \end{array}\right).            %,\\
%\Pi_{xx}=\frac{1}{2}\left(\begin{array}{ccc} -1 &
% 0 & 3\\ 0 & 2 & 0\\ 3 & 0 & -1
% \end{array}\right), ~~
%\Pi_{yy}=\frac{1}{2}\left(\begin{array}{ccc} -1 &
% 0 & -3\\ 0 & 2 & 0\\ -3 & 0 & -1
% \end{array}\right), ~~ \Pi_{zz}=\left(\begin{array}{ccc} 1 &
% 0 & 0\\ 0 & -2 & 0\\ 0 & 0 & 1
% \end{array}\right),\\
% \Pi_{xy}=\Pi_{yx}=i\frac{3}{2}\left(\begin{array}{ccc} 0 &
% 0 & -1\\ 0 & 0 & 0\\ 1 & 0 & 0
% \end{array}\right), ~~ \Pi_{xz}=\Pi_{zx}=\frac{3}{2\sqrt{2}}\left(\begin{array}{ccc} 0 &
% 1 & 0\\ 1 & 0 & -1\\ 0 & -1 & 0
% \end{array}\right), \\ \Pi_{yz}=\Pi_{zy}=i\frac{3}{2\sqrt{2}}\left(\begin{array}{ccc} 0 &
% -1 & 0\\ 1 & 0 & 1\\ 0 & -1 & 0
% \end{array}\right).
 \end{array}\label{eqsm}\end{equation}

Possibly, the nontrivial spin dynamics predicted in Refs.
\cite{Bar1,Bar3,Bar4} and conditioned by the tensor electric
polarizability of deuteron is the first example of importance of
spin tensor interactions in the physics of polarized beams. Tensor
interactions of deuteron can also be described with the method of
spin amplitudes. In this case, three-component spinors and
3$\times$3 matrices should be used. The method of spin amplitudes is
mathematically advantageous because transporting the three-component
spinor is much simpler than transporting the three-dimensional
polarization vector $\bm P$ and five independent components of the
polarization tensor $P_{ij}$ together.

We follow the traditional quantum mechanical approach perfectly
expounded by R. Feynman \cite{RF} and use the matrix Hamilton
equation and the matrix Hamiltonian $H$ for determining an evolution
of the spin wave function:
\begin{equation}
 \begin{array}{c}
 i \frac{d\Psi}{dt}=H\Psi, ~~~ \Psi=\left(\begin{array}{c}
C_{1}(t)
 \\ C_{0}(t) \\ C_{-1}(t) \end{array}\right),
 %~~ H=\left(\begin{array}{ccc} H_{11} & H_{12} & H_{13}\\
 % H_{21} & H_{22} & H_{23}\\
 % H_{31} & H_{32} & H_{33} \end{array}\right),
 ~~~ H_{ij}=<i|{\cal H}|j>,
 \end{array}\label{eq19t}\end{equation}
where $H$ is $3\times3$ matrix, $\Psi$ is the three-component spin
wave function (spinor), $H_{ij}=H_{ji}^\ast$ and $i,j=1,0,-1$. In
this equation, $H_{ij}$ are matrix elements of the Hamilton
operator ${\cal H}$.

A determination of spin dynamics can be divided into several
stages, namely

i) a solution of Hamilton equation (\ref{eq19t}) and a
determination of eigenvalues and eigenvectors of the Hamilton
matrix $H$;

ii) a derivation of spin wave function consisting in a solution of
a set of three linear algebraic equations;

iii) a calculation of time evolution of polarization vector and
polarization tensor.

\section{Hamilton Operator in a Cylindrical Coordinate System}

The spin dynamics can be analytically calculated when a storage
ring is either circular or divided into circular sectors by empty
spaces. In this case, the use of cylindrical coordinates can be
very successful. The particle spin motion in storage rings is
usually specified with respect to the particle trajectory. Main
fields are commonly defined relatively the cylindrical coordinate
axes. Therefore, the use of the cylindrical coordinates
considerably simplifies an analysis of spin rotation in the
horizontal plane (g$-$2 precession) and other spin effects.
Equation of spin motion in storage rings in a cylindrical
coordinate system has the form \cite{N84}
\begin{equation}
 \begin{array}{c}
\frac{d\bm S}{dt}=\bm\omega_a\times\bm S,~~~
\bm\omega_a=-\frac{e}{m}\left\{a\bm B-
 \frac{a\gamma}{\gamma+1}\bm\beta(\bm\beta\cdot\bm B)\right.%\nonumber\\
 +\left(\frac{1}{\gamma^2-1}-a\right)\left(\bm\beta\times\bm
 E\right)\\
 +\frac{1}{\gamma}\left[\bm B_\|
 -\frac{1}{\beta^2}\left(\bm\beta\times\bm
 E\right)_\|\right]%\nonumber\\
 \left.+ \frac{\eta}{2}\left(\bm
 E-\frac{\gamma}{\gamma+1}\bm\beta(\bm\beta\cdot\bm
 E)+\bm\beta\!\times\!\bm B\right)\!\right\},
 \end{array}\label{eq7}\end{equation}  where $a=(g-2)/2,~g=2\mu m/(eS)$,
$\eta=2dm/(eS)$, and $d$ is the EDM. The sign $\|$ means a
horizontal projection for any vector. In this work, we do not
consider effects caused by perturbations of particle
trajectory investigated in Ref. \cite{N84}. % and suppose $\bm\omega_a$ to
%be upward.
The quantity $\bm\omega_a$ is also equal to \cite{N84}
\begin{equation}
\bm\omega_a=\bm\Omega-\dot{\phi}\bm e_z,
\label{eqn5}\end{equation} where $\bm\Omega$ is the
Thomas-Bargmann-Michel-Telegdi (T-BMT) frequency \cite{BMT}
corrected for the EDM  \cite{N84,EDMNSPC,NPR,RPJ} and
$\dot{\phi}\bm e_z$ is the instantaneous angular frequency of
orbital revolution.

If we used the Hamiltonian of the particle with the EDM given in
the lab frame \cite{RPJ}, we would present the matrices $S_\rho$ and
$S_\phi$ in the form
\begin{equation} S_\rho=S_x\cos{\phi}+S_y\sin{\phi}, ~~~
S_\phi=-S_x\sin{\phi}+S_y\cos{\phi}. \label{eqsld}\end{equation}
However, this representation of spin matrices $S_\rho,S_\phi$
leads to cumbersome calculations because the azimuth $\phi$
defined by a particle position is time-dependent. Therefore, it is
helpful to consider spin effects in the frame rotating at the
instantaneous angular frequency of orbital revolution which is
almost equal to the cyclotron frequency. In this frame, the motion
of particles is relatively slow because it can be caused only by
beam oscillations and other deflections of particles from the
ideal trajectory. The equation of spin motion in the rotating
frame coincides with that in the cylindrical coordinate system
because the horizontal axis of this system rotates at the
instantaneous angular frequency of orbital revolution.

The Hamiltonians of the particle in the rotating frame and in the
lab one (${\cal H}$ and ${\cal H}_{lab}$, respectively) are
related by \cite{Mash1} \begin{equation} {\cal H}={\cal
H}_{lab}-\bm S\cdot\bm\omega,\label{eqMas}\end{equation} where
$\bm\omega$ is the observer's proper frequency of rotation
\cite{foo1}. In the considered case, this frequency coincides with
the instantaneous angular frequency of orbital revolution. The
relation between the Hamiltonian in the lab frame and the T-BMT
frequency corrected for the EDM is given by $${\cal H}_{lab}={\cal
H}_0+\bm S\cdot\bm\Omega,$$ where ${\cal H}_0$ is a sum of
spin-independent operators. Therefore, the Hamiltonian in the
rotating frame has the form
\begin{equation}
  {\cal H}={\cal H}_0+\bm S\cdot\bm\omega_a,
 \label{eq21}\end{equation} where $\bm\omega_a$ is defined by Eq.
 (\ref{eq7}). Evidently, Hamiltonian (\ref{eq21}) is consistent
 with Eq. (\ref{eq7}).

The particle in the rotating frame is localized and ideally is in
rest. Therefore, we can direct the $x$- and $y$-axes in this frame
along the radial and longitudinal axes, respectively. This
procedure is commonly used (see Refs. \cite{CR,MZ,MShY}) and
results in the direct substitution of spin matrices (\ref{eqsm})
for $S_\rho$ and $S_\phi$:
 \begin{equation}
\begin{array}{c}
S_\rho=S_x=\frac{1}{\sqrt{2}}\left(\begin{array}{ccc} 0 &
 1 & 0\\ 1 & 0 & 1\\ 0 & 1 & 0
 \end{array}\right), ~~ S_\phi=S_y=\frac{i}{\sqrt{2}}\left(\begin{array}{ccc} 0 &
 -1 & 0\\ 1 & 0 & -1\\ 0 & 1 & 0
 \end{array}\right).
 \end{array}\label{eq20}\end{equation} The matrix $S_z$ remains unchanged.
Similar substitution can be performed for the matrices $\Pi_{ij}$.

  The use of definition (\ref{eq20}) strongly simplifies calculations.

 If an interaction causes a correction to Hamilton operator
 (\ref{eq21}) and does not appreciably influence the particle motion,
 this correction is the same in the lab frame and the rotating
one.

In this work, we suppose $\bm B$ to be upward. For the deuteron,
$a<0$ and $(\omega_a)_z>0$.

\section{Corrections to the Hamilton Operator for Tensor Polarizabilities of Deuteron}

Corrections to the Hamilton operator for deuteron polarizabilities
contain scalar and tensor parts. The scalar part is
spin-independent and can be disregarded. General formulae used in
Refs. \cite{Bar1,Bar3,Bar4} are within first-order terms in the
normalized velocity $\beta$. In the present work, we derive exact
formulae for the configuration of main fields related to the
resonant deuteron EDM experiment (see Refs. \cite{YOr,YS}).
Because the Lorentz factor is planned to be $\gamma=1.28$
\cite{OMS} in this experiment, exact relativistic formulae are
needed.

Within first-order terms in $\beta$, the interaction Hamiltonian
depending on the electric and magnetic polarizabilities is given
by \begin{equation}
\begin{array}{c} V=V_e+V_m = -\frac{1}{2}\alpha_{ik}E'_iE'_k-
\frac{1}{2}\beta_{ik}B'_iB'_k,\\ \bm E'=\bm E+\bm\beta\times\bm B,
~~~ \bm B'=\bm B-\bm\beta\times\bm E,
\end{array}\label{eqwth}\end{equation} where $\alpha_{ik}$ and $\beta_{ik}$ are the tensors
of electric and magnetic polarizabilities, $\bm E'$ and $\bm B'$
are effective fields acting on a particle (fields in the
particle's rest frame, i.e., in the rotating frame). In this
approximation, the spin-dependent part of the Hamiltonian defined
by the tensor electric and magnetic polarizabilities is equal to
\cite{Bar1,Bar3,Bar4}
\begin{equation} \begin{array}{c}
V=-\alpha_T(\bm S\cdot\bm E')^2-\beta_T(\bm S\cdot\bm B')^2,
\end{array} \label{eqT} \end{equation}
where $\alpha_T$ and $\beta_T$ are the tensor electric and
magnetic polarizabilities, respectively.

The Baryshevsky effect takes place when one stimulates coherent longitudinal
oscillations of particles at a resonance frequency.
The angular frequency of forced longitudinal oscillations, $\omega$, is equal to
the difference between two radio frequencies (see Ref. \cite{OMS}). It should be very
close to the angular frequency of spin rotation (g$-$2 frequency),
$\omega_0$, and close to the eigenfrequency of free synchrotron
oscillations (synchrotron frequency) \cite{OMS}. The resonant
electric field in the particle's rest frame possesses the oscillating
longitudinal component $E'_\phi$ defined by the Lorentz
transformation of longitudinal electric field met by the oscillating particles
and the radial one $E'_\rho$ caused by the Lorentz
transformation of vertical magnetic field. The latter component
has a resonance part because of the modulation of the particle
velocity. In the present work, we consider only effects of
resonant fields on the BVP in ideal conditions and disregard
systematical errors listed in Section VII. Thus, we take into
consideration the constant vertical magnetic field and the
oscillating longitudinal electric one.

The relativistic formulae for the fields in the particle's rest
frame, $\bm E'$ and $\bm B'$, have the form
\begin{equation} \begin{array}{c}
\bm E'=\beta\gamma B_z\bm e_\rho+E_\phi\bm e_\phi, ~~~ \bm
B'=\gamma B_z\bm e_z.
\end{array} \label{eqTr} \end{equation}
The fields and the electromagnetic moments in the lab frame are
unprimed.

Induced electric and magnetic dipole moments in the particle's
rest frame caused by the tensor polarizabilities are equal to
\begin{equation} \begin{array}{c}
\bm d'=\alpha_T\{\bm S,(\bm S\cdot\bm E')\}, ~~~ \bm
m'=\beta_T\{\bm S,(\bm S\cdot\bm B')\},
\end{array} \label{eqTt} \end{equation}
where $\{\dots,\dots\}$ means an anticommutator.

The correction to the Hamilton operator for the tensor
polarizabilities of deuteron is equal to
\begin{equation} \begin{array}{c}
V=V_{lab}=-\frac12\left(\bm d\cdot\bm E+\bm m\cdot\bm B\right).
\end{array} \label{eqTrl} \end{equation}
This correction does not change the angular frequency of orbital
revolution. According to Eq. (\ref{eqMas}), the correction is the
same in the rotating frame and the lab one. Since the induced
dipole moments are proportional to the effective fields in the
particle's rest frame, the factor 1/2 appears.

To obtain the dipole moments in the lab frame, $\bm d$ and $\bm
m$, we can use the Hamilton operator for relativistic particles
with electric and magnetic dipole moments. For spin-1/2 particles,
it has been derived in Ref. \cite{RPJ}. The Hamilton operator for
spin-1 particles is similar, because it should be consistent with
the corresponding equation of spin motion (modified T-BMT
equation) which is valid for any spin.

If we neglect the normal magnetic moment $\mu_0=eS/m$ which is
small for nuclei, the relations between the electromagnetic
moments in two frames are given by
\begin{equation}
  \bm d=\bm d'-\frac{\gamma}{\gamma+1}\bm
  \beta(\bm d'\cdot\bm \beta)-\bm \beta\times\bm m', ~~~\bm m=\bm m'-\frac{\gamma}{\gamma+1}\bm
  \beta(\bm m'\cdot\bm \beta)+\bm
  \beta\times\bm d'.
\label{cl8}\end{equation} When $\bm d=e\bm l$, the relation
between $\bm d$ and $\bm d'$ arises from the Lorentz
transformation of the length of electric dipole, $\bm l$, because
the charge $e$ is a relativistic invariant.
Relations (\ref{cl8}) remain valid for induced electromagnetic
moments.

As a result, the correction to the Hamilton operator in the
rotating frame takes the form
\begin{equation} \begin{array}{c}
V=-\frac{1}{2\gamma}\left(\bm d'\cdot\bm E'+\bm m'\cdot\bm
B'\right)=-\frac{\alpha_T}{\gamma}(\bm S\cdot\bm
E')^2-\frac{\beta_T}{\gamma}(\bm S\cdot\bm B')^2.
\end{array} \label{eqTre} \end{equation}

Eq. (\ref{eqT}) is an approximate version of Eq. (\ref{eqTre}).

The equation of oscillatory motion of the particle has the form
\begin{equation}
\frac{d\bm p}{dt}=e\bm E.\label{eqo} \end{equation}

The quantity $\bm E$ in Eq. (\ref{eqo}) is the electric field met by
the particle. As a result of the coherent beam oscillations, this field
oscillates in the particle's rest frame. The angular frequency of velocity modulation
$\omega$ significantly differs from that of the resonator (see
Refs. \cite{YOr,OMS}) but should be very close to the angular frequency
of spin rotation $\omega_0$. The latter quantity is almost equal
to the vertical component of $\bm\omega_a$, because other
components of this vector are relatively small:
\begin{equation} \omega_0\equiv
\left(\omega_a\right)_z=-\frac{ea}{m}B_z.
\label{eqom}\end{equation} For the deuteron, $\omega_0>0$. %can be
%both positive and negative.

The modulation of normalized velocity can be given by (see Refs.
\cite{YOr,OMS})
\begin{equation}
\bm\beta=\frac{\bm p}{\sqrt{m^2+
p^2}}=\bm\beta_0+\Delta\beta_0\cdot\cos{(\omega t+\varphi)}\bm
e_\phi, \label{eq3}
\end{equation} where $$\bm\beta_0=\frac{\bm p_0}{\sqrt{m^2+
p_0^2}},~~~\gamma_0=\frac{\sqrt{m^2+p_0^2}}{m}.$$ Owing to this
modulation, the radial electric field in the particle's rest frame
has the oscillatory part. The effect of the modulation on the BVP
is described by the last term in Eq. (\ref{eq7}) proportional to
$\bm\beta\!\times\!\bm B$.

We can perform the calculation of electric field acting on the
particle to within first-order terms in $\Delta\beta_0$. The
particle momentum is defined by the equation
\begin{equation} \bm p=\frac{m\bm \beta}{\sqrt{1-
\beta^2}}=\bm p_0+\gamma_0^3 m\Delta\beta_0\cdot\cos{(\omega
t+\varphi)}\bm e_\phi. \label{eqnwl}\end{equation}

According to the result of differentiation of Eq. (\ref{eqnwl}) on
time, Eq. (\ref{eqo}) takes the form
\begin{equation}\bm E=-E_0\sin{(\omega t+\varphi)}\bm e_\phi,\label{eqf}\end{equation}
where
\begin{equation}
E_0=\frac{\gamma_0^3 m\omega}{e}\Delta\beta_0.
\label{eqE}\end{equation}

Eq. (\ref{eqTre}) can be transformed to the form
\begin{equation} \begin{array}{c}
V=-\frac{\alpha_T}{\gamma}(\beta\gamma B_z S_\rho+E_\phi
S_\phi)^2-\beta_T\gamma B^2_z S_z^2.
\end{array} \label{eqTrt} \end{equation}

An estimate of two terms in the formula for the effective electric
field $\bm E'$ [see Eq. (\ref{eqTr})] shows that the term
proportional to the magnetic field $B_z$ is much bigger for the
deuteron. To simplify the calculation, we neglect the effect of
the longitudinal electric field and use the approximation
\begin{equation} \begin{array}{c}
V=-\gamma B^2_z(\alpha_T\beta^2 S^2_\rho+\beta_T S_z^2).
\end{array} \label{eqapp} \end{equation}

The quantities $\gamma$ and $\beta^2\gamma$ can be expanded in
series of $\Delta\beta_0$:
\begin{equation}
 \begin{array}{c} \gamma=\gamma_0+\beta_0\gamma_0^3\cdot\Delta\beta_0\cos{(\omega
 t+\varphi)}+\frac14(1+3\beta_0^2\gamma_0^2)\gamma_0^3(\Delta\beta_0)^2
 \left\{1+\cos{\left[2(\omega t+\varphi)\right]}\right\},
 \\
 \beta^2\gamma=\beta_0^2\gamma_0+(2+\beta_0^2\gamma_0^2)\beta_0\gamma_0\cdot\Delta\beta_0\cos{(\omega
 t+\varphi)}\\
 +\frac14(2+5\beta_0^2\gamma_0^2
 +3\beta_0^4\gamma_0^4)\gamma_0(\Delta\beta_0)^2
 \left\{1+\cos{\left[2(\omega t+\varphi)\right]}\right\}.
 \end{array}\label{eqser}\end{equation}

Eqs. (\ref{eqapp}),(\ref{eqser}) define the corrections to the
Hamilton operator for the tensor polarizabilities of deuteron.

\section{Solution of Matrix Hamilton Equation}

Nonzero matrix elements of the spin operators contained by Eq.
(\ref{eqapp}) are
\begin{equation} \begin{array}{c}
(S_\rho^2)_{11}=(S_\rho^2)_{1,-1}=(S_\rho^2)_{-1,1}=(S_\rho^2)_{-1,-1}=\frac12,
~~~ (S_\rho^2)_{00}=1,~~~(S_z^2)_{11}=(S_z^2)_{-1,-1}=1.
\end{array} \label{eqTrm} \end{equation}

Therefore, matrix Hamiltonian (\ref{eq19t}) takes the form
\begin{equation}
 \begin{array}{c}
 H=\left(\begin{array}{ccc} E_{0}+\omega_0+{\cal A}+{\cal B} & 0 & {\cal A} \\
0 & E_{0}+2{\cal A} & 0 \\
{\cal A} & 0 & E_{0}-\omega_0+{\cal A}+{\cal B}
 \end{array}\right),
 \end{array}\label{eqMH}\end{equation}
where
\begin{equation}
 \begin{array}{c}
{\cal A}=a_0+a_1\cos{(\omega
 t+\varphi)}+a_2\cos{\left[2(\omega t+\varphi)\right]}, \\
{\cal B}=b_0+b_1\cos{(\omega
 t+\varphi)}+b_2\cos{\left[2(\omega t+\varphi)\right]}, \\
a_0=-\frac{1}{2}\alpha_TB_z^2\gamma_0\left[\beta_0^2+\frac{1}{4}(2+5\beta_0^2\gamma_0^2
 +3\beta_0^4\gamma_0^4)(\Delta\beta_0)^2\right],\\
 a_1=-\frac{1}{2}\alpha_TB_z^2(2+\beta_0^2\gamma_0^2)\beta_0\gamma_0\cdot\Delta\beta_0,\\
 a_2=-\frac{1}{8}\alpha_TB_z^2(2+5\beta_0^2\gamma_0^2
 +3\beta_0^4\gamma_0^4)\gamma_0(\Delta\beta_0)^2,\\
 b_0=-\beta_TB_z^2\gamma_0\left[1+\frac{1}{4}(1+3\beta_0^2\gamma_0^2)\gamma_0^2(\Delta\beta_0)^2\right],\\
 b_1=-\beta_TB_z^2\beta_0\gamma_0^3\cdot\Delta\beta_0,\\
 b_2=-\frac{1}{4}\beta_TB_z^2(1+3\beta_0^2\gamma_0^2)\gamma_0^3(\Delta\beta_0)^2,
 \end{array}\label{eqMHt}\end{equation}
and $E_0$ is the zero energy level.

In Hamiltonian (\ref{eqMH}), the EDM effect is not taken into
account. %consideration.

We consider the spin dynamics near a resonance. At the first
stage, it is useful to pass on to new amplitudes (see Ref.
\cite{RF}). This transformation brings real parts of diagonal
elements of the matrix Hamiltonian to zero. However, it does not
nullify the imagine parts of diagonal elements for unstable
particles (see Ref. \cite{CJP}). Evidently, Hamiltonian
(\ref{eqMH}) is real and the new amplitudes are equal to
\begin{equation}\begin{array}{c}
\gamma_1(t)=\exp{\left\{i\left[k_1t+\frac{a_1+b_1}{\omega}f(t)
% \right.\right.\\\left.\left.
 +\frac{a_2+b_2}{2\omega}g(t)\right]\right\}}C_1(t), \\
\gamma_0(t)=\exp{\left\{i\left[k_0t+\frac{2a_1}{\omega}f(t)
% \right.\right.\\\left.\left.
 +\frac{a_2}{\omega}g(t)\right]\right\}}C_0(t),
\\
\gamma_{-1}(t)=\exp{\left\{i\left[k_{-1}t+\frac{a_1+b_1}{\omega}f(t)
% \right.\right.\\\left.\left.
 +\frac{a_2+b_2}{2\omega}g(t)\right]\right\}}C_{-1}(t),\\
 k_1=E_{0}+\omega_0+a_0+b_0, ~~~ k_0=E_{0}+2a_0, ~~~
 k_{-1}=E_{0}-\omega_0+a_0+b_0,\\
 f(t)=\sin{(\omega
 t+\varphi)}-\sin{(\varphi)}, ~~~ g(t)=\sin{\left[2(\omega t+\varphi)\right]}
 -\sin{(2\varphi)}.
\end{array}\label{eq27u}\end{equation}

Dynamics of these amplitudes does not depend on the tensor
magnetic polarizability and is given by
\begin{equation}
\left\{\begin{array}{c} i\frac{d\gamma_1}{dt}={\cal
A}\exp{(2i\omega_0t)}\gamma_{-1}
\\ i\frac{d\gamma_0}{dt}=0\\
i\frac{d\gamma_{-1}}{dt}={\cal A}\exp{(-2i\omega_0t)}\gamma_1
\end{array}\right.. \label{eqgmu}\end{equation}
Eqs. (\ref{eq27u}),(\ref{eqgmu}) result in
\begin{equation}\begin{array}{c}
C_0(t)=\exp{\left\{-i\left[k_0t+\frac{2a_1}{\omega}f(t)
 +\frac{a_2}{\omega}g(t)\right]\right\}}C_0(0).
\end{array}\label{eq27}\end{equation}
Zero component of spin is not mixed with other components.

At the second stage, we can average over much longer time than the
oscillation period \cite{RF}. The relation $$\cos{(\zeta
 t+\eta)}=\frac12\left\{\exp{[i(\zeta
 t+\eta)]}+\exp{[-i(\zeta
 t+\eta)]}\right\}$$ can be used. There
are two resonant frequencies, $\omega=\omega_0$ and
$\omega=2\omega_0$. First of them corresponds to the resonance
condition in the deuteron EDM experiment \cite{YOr,YS}. We will
consider this case first.

When $\omega\approx\omega_0$, averaging over time results in
\begin{equation}
\left\{\begin{array}{c}
i\frac{d\gamma_1}{dt}=\frac{a_2}{2}\gamma_{-1}\exp{\left\{2i[(\omega_0-\omega)t-\varphi]\right\}}
\\ %i\frac{d\gamma_0}{dt}=0\\
i\frac{d\gamma_{-1}}{dt}=\frac{a_2}{2}\gamma_1\exp{\left\{-2i[(\omega_0-\omega)t-\varphi]\right\}}
\end{array}\right.. \label{eqavr}\end{equation}

At the third stage, we can use the following transformation:
\begin{equation}\begin{array}{c}
D_1(t)=\exp{\left[-i(\omega_0-\omega)t\right]}\gamma_1(t), \\
D_{-1}(t)=\exp{\left[i(\omega_0-\omega)t\right]}\gamma_1(t).
\end{array}\label{eq7u}\end{equation}

Transformed equation (\ref{eqavr}) can be written in the matrix
form:
\begin{equation}
 \begin{array}{c}
 i \frac{dD}{dt}=H'D, ~~~ H'=\left(\begin{array}{cc} \omega_0-\omega &
 \frac{a_2}{2}\exp{(-2i\varphi)}
 \\ \frac{a_2}{2}\exp{(2i\varphi)} & -(\omega_0-\omega) \end{array}\right),
 ~~~ D=\left(\begin{array}{c} D_{1}(t) \\ D_{-1}(t) \end{array}\right).
 \end{array}\label{eqnew}\end{equation}

Eq. (\ref{eqnew}) can be analytically solved. The solution of it
has the form
\begin{equation} \begin{array}{c}
D_1(t)=\left[\frac{\omega'+\omega_0-\omega}{2\omega'}D_1(0)+
\frac{{\cal E}}{2\omega'}D_{-1}(0)\right]\exp{(-i\omega't)}
\nonumber\\+\left[\frac{\omega'-(\omega_0-\omega)}{2\omega'}D_1(0)-
\frac{{\cal E}}{2\omega'}D_{-1}(0)\right]\exp{(i\omega't)},\nonumber\\
D_{-1}(t)=\left[\frac{{\cal
E}^\ast}{2\omega'}D_{1}(0)+\frac{\omega'-(\omega_0-\omega)}{2\omega'}D_{-1}(0)
\right]\exp{(-i\omega't)} \nonumber\\+\left[-\frac{{\cal
E}^\ast}{2\omega'}D_1(0)+\frac{\omega'+\omega_0-\omega}{2\omega'}D_{-1}(0)\right]\exp{(i\omega't)}
\end{array}
\label{eqin}
\end{equation}
or
\begin{equation} \begin{array}{c}
D_1(t)=\left[\cos{(\omega't)}-i
\frac{\omega_0\!-\!\omega}{\omega'}\sin{(\omega't)}
\right]D_1(0)-i\frac{{\cal
E}}{\omega'}\sin{(\omega't)}D_{-1}(0),\nonumber\\
D_{-1}(t)=-i\frac{{\cal
E}^\ast}{\omega'}\sin{(\omega't)}D_1(0)+\left[\cos{(\omega't)}+i
\frac{\omega_0\!-\!\omega}{\omega'}\sin{(\omega't)}
\right]D_{-1}(0),
\end{array}
\label{eq14}
\end{equation}
where %the angular frequency of spin oscillation is \cite{FF,MSYS}}
\begin{equation}\omega'=\sqrt{(\omega_0-\omega)^2+{\cal E}{\cal E}^\ast},
~~~ {\cal E}=\frac{a_2}{2}\exp{(-2i\varphi)}.\label{eq15}
\end{equation}
The angular frequency of spin oscillation is equal to $2\omega'$.

The initial spin amplitudes take the form
\begin{equation} \begin{array}{c}
  C_1(t)=\exp{\left\{-i\left[(E_{0}+\omega+a_0+b_0)t+\frac{a_1+b_1}{\omega}f(t)
 +\frac{a_2+b_2}{2\omega}g(t)\right]\right\}}D_1(t),\\
 C_{-1}(t)=\exp{\left\{-i\left[(E_{0}-\omega+a_0+b_0)t+\frac{a_1+b_1}{\omega}f(t)
 +\frac{a_2+b_2}{2\omega}g(t)\right]\right\}}D_{-1}(t),\\ C_1(0)=D_1(0),
 ~~~C_{-1}(0)=D_{-1}(0).
\end{array}\label{eq12}\end{equation}

The resonance at the doubled frequency $\omega\approx2\omega_0$
can be investigated in a similar way. The evolution of the spin
amplitudes is given by
\begin{equation} \begin{array}{c}
  C_1(t)=\exp{\left\{-i\left[(E_{0}+\frac{\omega}{2}+a_0+b_0)t+\frac{a_1+b_1}{\omega}f(t)
 +\frac{a_2+b_2}{2\omega}g(t)\right]\right\}}D_1(t),\\
 C_0(t)=\exp{\left\{-i\left[(E_{0}+2a_0)t+\frac{2a_1}{\omega}f(t)
 +\frac{a_2}{\omega}g(t)\right]\right\}}C_0(0),\\
 C_{-1}(t)=\exp{\left\{-i\left[(E_{0}-\frac{\omega}{2}+a_0+b_0)t+\frac{a_1+b_1}{\omega}f(t)
 +\frac{a_2+b_2}{2\omega}g(t)\right]\right\}}D_{-1}(t),\\ C_1(0)=D_1(0),
 ~~~C_{-1}(0)=D_{-1}(0),
\end{array}\label{eq12t}\end{equation}
where
\begin{equation} \begin{array}{c}
D_1(t)=\left(\cos{\frac{\omega''t}{2}}-i
\frac{2\omega_0\!-\!\omega}{\omega''}\sin{\frac{\omega''t}{2}}
\right)D_1(0)-i\frac{2{\cal
E}'}{\omega''}\sin{\frac{\omega''t}{2}}D_{-1}(0),\nonumber\\
D_{-1}(t)=-i\frac{2{{\cal
E}'}^\ast}{\omega''}\sin{\frac{\omega''t}{2}}D_1(0)+\left(\cos{\frac{\omega''t}{2}}+i
\frac{2\omega_0\!-\!\omega}{\omega''}\sin{\frac{\omega''t}{2}}
\right)D_{-1}(0),\\
{\cal E}'=\frac{a_1}{2}\exp{(-i\varphi)},
\end{array}
\label{eq14t}
\end{equation}
and the angular frequency of spin oscillation is equal to
\begin{equation}\omega''=\sqrt{(2\omega_0-\omega)^2+4{\cal E}'{{\cal E}'}^\ast}.
\label{eq15t}\end{equation}

\section{Spin Dynamics Caused by Tensor Polarizabilities of Deuteron}

For spin-1 particles, three components of polarization vector and
related components of polarization tensor are defined by
\begin{equation}
 \begin{array}{c}
P_\rho=\frac{1}{\sqrt2}(C_1C_0^\ast+C_1^\ast C_0+C_0C_{-1}^\ast+C_0^\ast C_{-1}),\\
P_\phi=\frac{i}{\sqrt2}(C_1C_0^\ast-C_1^\ast C_0+C_0C_{-1}^\ast-C_0^\ast C_{-1}),\\
P_z=(C_1C_1^\ast-C_{-1}C_{-1}^\ast),\\
P_{\rho\rho}=\frac{3}{2}(C_1C_{-1}^\ast+C_1^\ast C_{-1}+C_0C_0^\ast)-\frac{1}{2},\\
P_{\phi\phi}=-\frac{3}{2}(C_1C_{-1}^\ast+C_1^\ast
C_{-1}-C_0C_0^\ast)-\frac{1}{2},\\
P_{\rho\phi}=i\frac{3}{2}(C_1C_{-1}^\ast-C_1^\ast C_{-1}).
\end{array}
\label{eqpvu}
\end{equation}

The horizontal components, $P_\rho$ and $P_\phi$, do not give
necessary information about the investigated effect because they
undergo fast oscillations caused by the g$-$2 spin precession. The
change of the vertical component, $P_z$, is a relatively slow
process.

The quantity $P_z$ does not depend on $C_0$. Since
$C_1C_1^\ast=D_1D_1^\ast,~C_{-1}C_{-1}^\ast=D_{-1}D_{-1}^\ast$,
the BVP is caused by the tensor electric polarizability and is not
affected by the tensor magnetic one. However, this conclusion is
not valid if the deuteron possesses the EDM. In this case, the
tensor magnetic polarizability leads to splitting of resonance
frequency \cite{Bar1,Bar3,Bar4}.

When $\omega\approx\omega_0$, the evolution of the vertical
component of polarization vector is expressed by
\begin{equation}\begin{array}{c}
P_z(t)=\left[1-\frac{{\cal
E}_0^2}{{\omega'}^2}\left(1-\cos{(2\omega't)}\right)\right]P_z(0)\\+\frac{2{\cal
E}_0}{3\omega'}\left\{\frac12[P_{\rho\rho}(0)-P_{\phi\phi}(0)]\left[\frac{\omega_0-\omega}{\omega'}
\cos{(2\varphi)}\left(1-\cos{(2\omega't)}\right)-\sin{(2\varphi)}\sin{(2\omega't)}\right]\right.
\\\left.+ P_{\rho\phi}(0)\left[\frac{\omega_0-\omega}{\omega'}
\sin{(2\varphi)}\left(1-\cos{(2\omega't)}\right)+\cos{(2\varphi)}\sin{(2\omega't)}\right]\right\},
~~~ {\cal E}_0=\frac{a_2}{2},
\end{array} \label{eq2} \end{equation}
where the quantities $a_2$ and $\omega'$ are defined by Eqs.
(\ref{eqMHt}) and (\ref{eq15}), respectively.

For a vector-polarized deuteron beam, the related components of
polarization vector and polarization tensor have the form
\begin{equation}\begin{array}{c}
P_{z}=\cos{(\theta)}, ~~~ P_{\rho\rho}=\frac12\left[3\sin^2{(\theta)}\cos^2{(\psi)}-1\right],\\
%~~~
P_{\phi\phi}=\frac12\left[3\sin^2{(\theta)}\sin^2{(\psi)}-1\right],
~~~ P_{\rho\phi}=\frac34\sin^2{(\theta)}\sin{(2\psi)},
\end{array} \label{eq4} \end{equation}
where $\theta$ and $\psi$ are the spherical angles defining the
spin direction in the rotating frame. The azimuth $\psi=0$ characterizes the spin %When $\theta=\pi/2$,
directed radially outward.

The projection of deuteron spin onto the preferential direction
can be equal to zero. The beam possessing such a polarization is
tensor polarized. The vector polarization of this beam is zero.
The components of polarization vector and polarization tensor take
the form
\begin{equation}\begin{array}{c}
P_{\rho}=P_{\phi}=P_{z}=0, ~~~ P_{\rho\rho}=-3\sin^2{(\theta)}\cos^2{(\psi)}+1,\\
%~~~
P_{\phi\phi}=-3\sin^2{(\theta)}\sin^2{(\psi)}+1, ~~~
P_{\rho\phi}=-\frac32\sin^2{(\theta)}\sin{(2\psi)},
\end{array} \label{eqts} \end{equation}
where $\theta$ and $\psi$ are the spherical angles stated above.
When the polarization of deuteron beam is vector, Eqs.
(\ref{eq2}),(\ref{eq4}) result in
\begin{equation}\begin{array}{c}
P_z(t)=\left[1-\frac{{\cal
E}_0^2}{{\omega'}^2}\left(1-\cos{(2\omega't)}\right)\right]\cos{(\theta)}\\+\frac{{\cal
E}_0}{2\omega'}\sin^2{(\theta)}\left\{\frac{\omega_0-\omega}{\omega'}
\cos{\left[2(\psi-\varphi)\right]}\left[1-\cos{(2\omega't)}\right]+\sin{\left[2(\psi-\varphi)\right]}
\sin{(2\omega't)}\right\}.
\end{array} \label{eq5} \end{equation}

If the resonance is perfect, the initial beam polarization is
horizontal $[P_z(0)=0]$, and $\omega't<<0$, Eq. (\ref{eq2}) takes
the form
\begin{equation}\begin{array}{c}
P_z(t)=\frac{2}{3}a_2t\left[P_{\rho\phi}(0)\cos{(2\varphi)}-\frac{P_{\rho\rho}(0)
-P_{\phi\phi}(0)}{2}\sin{(2\varphi)}\right].
\end{array} \label{eqr} \end{equation}
In this case, the vertical spin component grows linearly with time
and
\begin{equation}\begin{array}{c}
\frac{dP_z}{dt}=\frac{2}{3}a_2\left[P_{\rho\phi}(0)\cos{(2\varphi)}-\frac{P_{\rho\rho}(0)
-P_{\phi\phi}(0)}{2}\sin{(2\varphi)}\right].
\end{array} \label{eqrg} \end{equation}

For the resonance at the doubled frequency
$\omega\approx2\omega_0$, the evolution of the vertical component
of polarization vector is given by
\begin{equation}\begin{array}{c}
P_z(t)=\left[1-\frac{4{{\cal
E}'}_0^2}{{\omega''}^2}\left(1-\cos{(\omega''t)}\right)\right]P_z(0)\\+\frac{2{\cal
E}'_0}{3\omega''}\left\{[P_{\rho\rho}(0)-P_{\phi\phi}(0)]\left[\frac{2\omega_0-\omega}{\omega''}
\cos{(\varphi)}\left(1-\cos{(\omega''t)}\right)-\sin{(\varphi)}\sin{(\omega''t)}\right]\right.
\\\left.+ 2P_{\rho\phi}(0)\left[\frac{2\omega_0-\omega}{\omega''}
\sin{(\varphi)}\left(1-\cos{(\omega''t)}\right)+\cos{(\varphi)}\sin{(\omega''t)}\right]\right\},
~~~ {\cal E}'_0=\frac{a_1}{2},
\end{array} \label{eq2d} \end{equation}
where the quantities $a_1$ and $\omega''$ are defined by Eqs.
(\ref{eqMHt}) and (\ref{eq15t}), respectively.

If the resonance is perfect, the initial beam polarization is
horizontal, and $\omega''t<<0$, Eq. (\ref{eq2d}) takes the form
\begin{equation}\begin{array}{c}
P_z(t)=\frac{2}{3}a_1t\left[P_{\rho\phi}(0)\cos{(\varphi)}-\frac{P_{\rho\rho}(0)
-P_{\phi\phi}(0)}{2}\sin{(\varphi)}\right].
\end{array} \label{eqrd} \end{equation}

Eqs. (\ref{eqMHt}),(\ref{eqr}), and (\ref{eqrd}) show that
resonance frequency doubling leads to a dramatic amplification of
the Baryshevsky effect. When the frequency is doubled, the EDM
effect becomes nonresonant. In this case, it does not influence
the spin dynamics.

\section{Measurement of Tensor Electric
Polarizability of Deuteron in Storage Ring Experiments}

To discover the Baryshevsky effect, it is necessary to stimulate
the BVP conditioned by the tensor electric polarizability of
deuteron and to avoid a similar effect caused by the magnetic
moment. It is known that the magnetic resonance takes place when
the particle placed in a uniform vertical magnetic field is also
affected by a horizontal magnetic field oscillating at a frequency
close to the frequency of spin rotation (see, e.g., Ref.
\cite{Slic}). The magnetic resonance results in the BVP for a
horizontally polarized beam.

Evidently, the magnetic resonance cannot take place when the
electric field is longitudinal, because nothing but
the oscillating electric field appears in the particle's rest
frame. Since the frequencies of betatron oscillations are chosen
to be far from resonances, these oscillations cannot lead to the
resonance effect. However, the resonance is caused by the tensor
electric polarizability of deuteron. %The oscillating electric
%field stimulates the coherent longitudinal oscillations of particles.
The electric field in the particle's rest frame
possesses the oscillating longitudinal component $E'_\phi$
and the radial one $E'_\rho$ caused by
the Lorentz transformation of the vertical magnetic field. The
latter component has a resonance part because of the modulation of
the particle velocity.

To measure the effect, some resonators (rf cavities) should be
used. The electric field in a resonator is generated along the
central line and the magnetic field is orthogonally directed
\cite{JJ}. The magnetic field along the central line is equal to
zero. If the rf cavities are perfectly placed and longitudinally
directed, the magnetic field cannot stimulate any resonance
effect. In this case, the observed BVP corresponds to the definite
value of the tensor electric polarizability.

However, both a displacement and an angular deviation of the
center line of the rf cavities away from an average particle
trajectory lead to a similar behavior of spin imitating the
Baryshevsky effect. As a result, they create
systematical errors in measurement of the tensor electric polarizability. %\cite{YS,YSNotes}.
Most of these errors are not in resonance with the spin precession
in the horizontal plane. Therefore, they create background and
result in fast oscillations of the vertical component of
polarization vector (see Refs. \cite{YOr,YS,OMS,YSNotes}). Besides
this effect, the systematical error can be caused by a radial
magnetic field in the particle's rest frame oscillating at the
resonant frequency. In the deuteron EDM experiment, a similar
error will be eliminated by alternately producing two sub-beams
with different betatron tunes \cite{YOr,OMS,YSNotes}. In the
deuteron tensor-electric-polarizability one, the resonant radial
magnetic field in the particle's rest frame is much less important
when a tensor polarized beam is used (see below). We calculate
only effects of resonant fields on the BVP in ideal conditions and
disregard systematical errors. Thus, we take into consideration
the constant vertical magnetic field and the oscillating
longitudinal electric one.

The measurement of the tensor electric polarizability of deuteron in a
storage ring needs the field configuration similar to that
proposed for the deuteron EDM experiment \cite{YOr,YS,OMS}.
However, the resonance frequency should be doubled
($\omega\approx2\omega_0$). Resonance frequency doubling cannot be
implemented in the designed EDM ring. In this ring, the
eigenfrequency of free synchrotron oscillations must be chosen
close to the g$-$2 frequency, $\omega_a$, and the resonance effect
is created by the beatings between two rf frequencies \cite{OMS}.
Therefore, the measurement of the tensor electric polarizability of
deuteron needs another ring or at least rf cavities different from
that developed for the deuteron EDM experiment.

However, the Baryshevsky effect caused by the tensor electric
polarizability of deuteron should be taken into account when
performing the deuteron EDM experiment \cite{Bar1,Bar3,Bar4}. This
effect results in the similar BVP and can imitate the presence of
the deuteron EDM of order of $d\sim10^{-29}$ e$\cdot$cm. An
attainment of such an accuracy is the goal of the storage ring EDM
experiment \cite{YOr,YS,OMS}.

The EDM-dependent evolution of deuteron spin in this experiment
has been calculated in detail in Ref. \cite{lanl}. The dynamics of
the vertical component of polarization vector is given by
\begin{equation}
\begin{array}{c}
P_z^{(EDM)}(t)=\frac{{\cal
E}_0''}{\Omega'}\left\{\frac{\omega_0-\omega}{\Omega'}
\cos{(\psi-\varphi)}\left[1-\cos{(\Omega't)}\right]+\sin{(\psi-\varphi)}\sin{(\Omega't)}\right\},
\end{array}\label{eqhnu}\end{equation}
where
\begin{equation}\begin{array}{c}
\Omega'=|\bm\Omega'|=\sqrt{(\omega_0-\omega)^2+{{\cal E}_0''}^2},
\end{array}\label{eqft}
\end{equation}
\begin{equation}\begin{array}{c}
{\cal
E}_0''=-\frac{1}{2}dB_z\cdot\Delta\beta_0\left(1+\frac{a\gamma_0^2\omega}{\omega_0}\right),
\end{array}\label{eqfn}
\end{equation}
and the azimuth $\psi$ defines the direction of spin at zero time.
The initial polarization is supposed to be horizontal.

When $\Omega't\ll1$,
\begin{equation}\begin{array}{c}
P_z^{(EDM)}={\cal E}_0''t\sin{(\psi-\varphi)}
\\=
-\frac{1}{2}dB_z\Delta\beta_0\left(1+\frac{a\gamma_0^2\omega}{\omega_0}\right)t\sin{(\psi-\varphi)}.
\end{array}\label{eqfa}
\end{equation}

We can evaluate the expected sensitivity in the measurement of the
tensor electric polarizability of deuteron with the comparison of
Eqs. (\ref{eq2d}) and (\ref{eqhnu}) and the use of Eq. (\ref{eq4})
and the sensitivity of the deuteron EDM experiment estimated in
Ref. \cite{OMS}. For the deuteron, $a=a_d=-0.14299$.
The sensitivity to the EDM of $d=1\times10^{-29}$ e$\cdot$cm
corresponds to the accuracy of $\delta\alpha_T=1.2\times10^{-43}$ %the accuracy of $\delta\alpha_T=1.7417\times10^{-43}$
cm$^3$ when $\omega\approx2\omega_0$ and Eqs.
(\ref{eq2d}),(\ref{eqhnu})--(\ref{eqfn}) are used. This estimate
is based on the values of $\gamma_0=1.28,~\beta_0=0.625,~\Delta
v_0=3.5\times10^6$ m/s, and $B_z=3$ T \cite{OMS}.
%$\gamma_0=1.2805,~\beta_0=0.6246,~\Delta\beta_0/\beta_0=0.01$, and $B_z=2$ T \cite{YKS}.
There are three independent theoretical predictions for the value
of the tensor electric polarizability of deuteron, namely
$\alpha_T=-6.2\times10^{-41}$ cm$^3$ \cite{CGS},
$-6.8\times10^{-41}$ cm$^3$ \cite{JL},  and $3.2\times10^{-41}$
cm$^3$ \cite{FP}. Two first values are very close to each other
but they do not agree with the last result.

In all probability, the best sensitivity in the measurement of
$\alpha_T$ can be achieved with the use of a tensor polarized
deuteron beam. The initial preferential direction of deuteron
polarization should be horizontal. When the vector polarization of
such a beam is zero, any spin rotation does not occur. In this
case, there are no related systematical errors caused by the
radial magnetic field and some other reasons. In the general case,
such systematical errors are proportional to a residual vector
polarization of the beam. This advantage leads to a sufficient
increase in experimental accuracy. When $\omega\approx2\omega_0$,
the equation describing the evolution of beam polarization takes
the form
\begin{equation}\begin{array}{c}
P_z(t)=-\frac{2{\cal
E}'_0}{\omega''}\sin^2{(\theta)}\Biggl\{\frac{2\omega_0-\omega}{\omega''}
\cos{(2\psi-\varphi)}\left[1-\cos{(\omega''t)}\right]
\\%\right.\\\left.
+\sin{(2\psi-\varphi)}\sin{(\omega''t)}\Biggr\}.
\end{array} \label{eq2dp} \end{equation}

In this case, the preliminary estimate of experimental accuracy is
$\delta\alpha_T\sim10^{-45}\div10^{-44}$ cm$^3$.

When $\theta=\pi/2$, the natural choice of phase
$$\varphi=2\psi\pm\frac{\pi}{2}$$
brings Eq. (\ref{eq2dp}) to the form
\begin{equation}\begin{array}{c}
P_z(t)=\pm\frac{2{\cal E}'_0}{\omega''}
\sin{\left(\omega''t\right)}.
\end{array} \label{eq10} \end{equation}

Other possibilities, $\varphi=2\psi$ and $\varphi=2\psi\pm\pi$,
lead to the equation
\begin{equation}\begin{array}{c}
P_z(t)=\pm\frac{4{\cal E}'_0(2\omega_0-\omega)}{{\omega''}^2}
\sin^2{\left(\frac{\omega''t}{2}\right)}.
\end{array} \label{eqot} \end{equation}
The dependence of $P_z(t)$ on time becomes quadratic when
$\omega''t\ll1$. Therefore, these possibilities are less useful.
However, they can be used for checking the result.

The deuteron tensor-electric-polarizability experiment essentially
differs from the deuteron EDM one by a nonnecessity of careful
checking systematical errors caused by the horizontal magnetic
field in the particle's rest frame. The use of a tensor polarized
beam makes it possible to avoid any spin rotations and to cancel
all related systematical errors. The BVP of such a beam is defined
only by the tensor electric polarizability of deuteron. It is a
great advantage because the elimination of similar systematical
errors is one of main problems for the deuteron EDM experiment
\cite{OMS,YSNotes}. A residual vector polarization of the beam
together with a resonant magnetic field in the particle's rest
frame can result in a false signal. However, a necessary
correction into the BVP can be made with a longitudinally
vector-polarized beam. It is important that the BVP caused by any
systematical error is overturned and the BVP defined by the tensor
electric polarizability remains the same when reversing the
polarization of this beam [see Eq. (\ref{eq5})]. This property
brings a easy differentiation between the Baryshevsky effect and
false signals for vector-polarized beams. In all probability, the
deuteron tensor-electric-polarizability experiment can be made at
one of existing rings.

\section{Differentiation of Effects of EDM and Tensor Electric
Polarizability in the Deuteron EDM Experiment}

Eqs. (\ref{eq2}) and (\ref{eqhnu})--(\ref{eqfn}) describing the
effects of the tensor electric polarizability and the EDM on the
spin dynamics in the EDM experiment essentially differ. Therefore,
these effects can be differentiated.

When the initial polarization of deuteron beam is horizontal, Eq.
(\ref{eq5}) takes the form
\begin{equation}\begin{array}{c}
P_z^{(tensor)}(t)=\frac{{\cal
E}_0}{2\omega'}\left\{\frac{\omega_0-\omega}{\omega'}
\cos{[2(\psi-\varphi)]}\left[1-\cos{(2\omega't)}\right]+\sin{[2(\psi-\varphi)]}\sin{(2\omega't)}\right\}.
\end{array} \label{eq2h} \end{equation}

For the EDM experiment, the choice of phase
$$\varphi=\psi\pm\frac{\pi}{2}$$
is necessary. This choice results in
\begin{equation}
\begin{array}{c}
P_z^{(EDM)}(t)=\pm\frac{{\cal
E}_0''}{\Omega'}\sin{(\Omega't)},~~~%\\
P_z^{(tensor)}(t)=\pm\frac{{\cal
E}_0(\omega_0-\omega)}{{\omega'}^2}\sin^2{(\omega't)}.
\end{array}\label{eqjf}\end{equation}

Since the quantities ${\cal E}_0'',{\cal E}_0$ are very small,
$\Omega'\approx\omega'\approx|\omega_0-\omega|$. When we analyze
only the ratio of amplitudes in Eq. (\ref{eqjf}) which is
approximately equal to $2{\cal E}_0''/{\cal E}_0$, we obtain that
the values of $\alpha_T$ found in Refs. \cite{CGS,JL,FP}
correspond to the false EDM moments of
$|d|=3\times10^{-29},~3\times10^{-29},$ and $2\times10^{-29}$
e$\cdot$cm, respectively. However, the EDM contribution to $P_z$
grows linearly with time when $\Omega't\ll1$, while the
tensor-electric-polarizability contribution is negligible in this
case. Therefore, keeping the frequency and phase of the coherent
longitudinal oscillations almost equal to the frequency and phase of the
spin rotations makes it possible to cancel the effect of the
tensor electric polarizability in the framework of the deuteron
EDM experiment. The same conclusion was recently drawn in Ref. \cite{Orlov}.
Nevertheless, the tensor electric polarizability of deuteron should
be taken into account. To check the possible existence of the
deuteron EDM, one can also use other possibilities of separating
the EDM and Baryshevsky effects listed below.

1) The spin dynamics caused by first-order interactions (including
the EDM effect) and second-order interactions (including the
Baryshevsky effect) is defined by the operator equations of spin
motion
\begin{equation}\frac{d\bm S}{dt}=A\bm\Omega\times\bm S
\label{eqfo}
\end{equation}
and
\begin{equation}\frac{dS_i}{dt}=\beta_{ijk}S_jS_k,
\label{eqfq}
\end{equation}
respectively. Therefore, the EDM effect reverses the sign when the
beam polarization is reversed while the sign of the Baryshevsky
effect remains unchanged.

2) Since both the EDM and Baryshevsky effects depend on the
difference $\psi-\varphi$, reversing the beam polarization
($\psi\rightarrow\psi+\pi$) is equivalent to the transition to the
opposite phase ($\varphi\rightarrow\varphi+\pi$). Naturally, such
a transition is technically simpler.

If two measurements fulfilled according to points 1) or 2) give
the values $P_{z1}$ and $P_{z2}$ for the BVP, the EDM and
Baryshevsky effects are characterized by the values
$(P_{z1}-P_{z2})/2$ and $(P_{z1}+P_{z2})/2$, respectively.

3) In the particle rest frame, the EDM and Baryshevsky effects are
linear and quadratic in the electric field, respectively. The
experimental dependence can be determined with changing the
amplitude of the field in the resonators.

4) The frequency of BVP caused by the Baryshevsky effect is
approximately twice as large as that conditioned by the EDM.

5) The use of tensor polarized deuteron beam even at the angular
frequency $\omega\approx\omega_0$ cancels the EDM effect and main
systematical errors. The evolution of beam polarization is given
by
\begin{equation}
\begin{array}{c}
P_z^{(tensor)}(t)=-\frac{a_2}{2\omega'}\left\{\frac{\omega_0-\omega}{\omega'}
\cos{[2(\psi-\varphi)]}\left[1-\cos{(2\omega't)}\right]+\sin{[2(\psi-\varphi)]}
\sin{(2\omega't)}\right\},
\end{array}\label{eqjt}\end{equation}
if the initial polarization is defined by Eq. (\ref{eqts}).

Thus, the EDM and Baryshevsky effects can be effectively
differentiated and the latter effect can be cancelled in the
framework of the deuteron EDM experiment.

\section{Discussion and Summary}

The spin dynamics conditioned by the tensor electric
polarizability of deuteron has been calculated for the first time
in Refs. \cite{Bar1,Bar3,Bar4}. To compare our results with those
obtained in \cite{Bar1,Bar3,Bar4}, it is helpful to introduce the
effective field defined by
\begin{equation}
\begin{array}{c}
E_{eff}^{2}=\beta^2\gamma B_z^2.
%, ~~~E_{eff}=E_{eff}^{(0)}+\delta E_{eff}^{(0)}\cos{(\omega t+\varphi)}.
%, ~~~ E_{eff}^{(0)}=\beta_0^2\gamma_0 B_z^2.
\end{array}\label{efff}\end{equation}

In Refs. \cite{Bar1,Bar3,Bar4}, the %relativistic
effect has been described to within first-order terms in $\beta$.
In this approximation, the squared effective field is equal to
\begin{equation}
\begin{array}{c}
E_{eff}^{2}=\left(E_{eff}^{(0)}\right)^2+2B_z^2\beta_0\cdot\Delta\beta_0\cos{(\omega
t+\varphi)}+\frac12B_z^2(\Delta\beta_0)^2\cos{[2(\omega
t+\varphi)]}.
\end{array}\label{efq}\end{equation}

The evolution of polarization vector is given by Eqs. (24),(29) in
Ref. \cite{Bar3} and Eqs. (44),(49) in Ref. \cite{Bar4}. The final
equation has the form
\begin{equation}\begin{array}{c}
\frac{dP_z}{dt}=-\frac12\Delta
\Omega_T\cos{(2\Omega_ft+2\varphi_f)}\left[P_{\rho\phi}(0)\cos{(2\Omega
t)}-\frac{P_{\rho\rho}(0)-P_{\phi\phi}(0)}{2}\sin{(2\Omega
t)}\right],
\end{array} \label{eqbar} \end{equation}
where
\begin{equation}\Delta\Omega_T=-\frac23\alpha_T B_z^2(\Delta\beta_0)^2,\label{eqdel} \end{equation}
$\Omega$ corresponds to our designation $\omega_0$, and
$\varphi_f=\varphi+\pi/2$. The spin rotation is supposed to be
clockwise in Refs. \cite{Bar1,Bar3,Bar4} and counter-clockwise
($\omega_0>0$) in the present work.  Therefore,
$\Omega_f\approx-\Omega$ and averaging Eq. (\ref{eqbar}) over time
with allowance for Eq. (\ref{eqdel}) results in
\begin{equation}\begin{array}{c}
\frac{dP_z}{dt}=-\frac{1}{12}\alpha_T
B_z^2(\Delta\beta_0)^2\left\{2P_{\rho\phi}(0)\cos{(2\varphi)}-[P_{\rho\rho}(0)
-P_{\phi\phi}(0)]\sin{(2\varphi)}\right\}.
\end{array} \label{eqb} \end{equation}
This equation fully agrees with Eq. (\ref{eqrg}).

Agreement of the results obtained in Refs. \cite{Bar3,Bar4} and in
the present work confirms their validity. The method used in Refs.
\cite{Bar3,Bar4} is less convenient to calculate the spin dynamics
in oscillatory external fields than in static ones. In the theory
of magnetic resonance, the transition to a rotating frame is
commonly used \cite{Slic}. Evidently, the transition to the
rotating frame is necessary in order to determine the spin
dynamics with the method developed in Refs. \cite{Bar3,Bar4} when
$\omega\neq\omega_0~(\Omega\neq|\Omega_f|)$.

The calculated effect of the BVP caused by the tensor electric
polarizability of deuteron is an exciting example of new spin
physics brought by tensor interactions. In the considered case,
the deuteron spin is governed by the electromagnetic interaction.
The similar effect investigated in works by V. Baryshevsky
\cite{VG,VGB} is affected by the strong interaction of deuteron
with nuclear matter. These effects stimulated by tensor
interactions result in the transformation of tensor polarization
into the vector one and the other way round.

Eqs. (\ref{eqin}) shows that the spin-up state is converted into
the spin-down state and the other way round. This property is
caused by the nondiagonal terms in Hamiltonian (\ref{eqnew}). As a
result, the vertical component of polarization vector oscillates.
This phenomenon is similar to light birefringence in crystals
\cite{VG,VGB}.

A similar behavior of spin takes place at the NMR when a nucleus
is placed into a resonant horizontal magnetic field. As is well
known, the NMR also consists in an oscillation of $P_z$. However,
there is an essential difference between two effects. The
Baryshevsky effect exists even when the beam is tensor polarized,
while the NMR does not change the beam polarization in this case.

The calculation shows that the Baryshevsky effect can %easily
be observed in storage rings. Performing the measurements with the
use of resonance $\omega\approx2\omega_0$ offers an opportunity to
measure the deuteron's tensor electric polarizability with the
accuracy of $10^{-45}\div10^{-44}$ cm$^3$ ($10^{-6}\div10^{-5}$
fm$^3$).

It is also possible to use low-energy deuterons in a Penning trap.

The problem of influence of the tensor electric polarizability on spin
dynamics in the deuteron EDM experiment in storage rings has been
investigated. The EDM and Baryshevsky effects can be effectively
differentiated and the latter effect can be cancelled in the
framework of this experiment.

In the present work, general formulae describing the BVP
conditioned by the tensor electric polarizability have been
derived. Calculated formulae agree with the previous results
\cite{Bar1,Bar3,Bar4} obtained in a more particular case. The
method based on the use of Hamiltonian approach and spin wave
functions happens to be very convenient for the investigation of
the effect.

\section*{Acknowledgements}

The author would like to thank V.G. Baryshevsky for helpful
discussions and acknowledge a financial support by BRFFR (grant
No. $\Phi$06D-002).


\begin{thebibliography}{}
\bibitem{foo1}
The observer's velocity in the lab frame is supposed to be zero,
because just this case corresponds to the real experimental
situation.
\end{thebibliography}
\end{document}